\documentclass[aps,prb,floatfix,superscriptaddress,twocolumn,floats]{revtex4}
\include{graphicx}

\begin{document}

\title{A Classical Picture of the Role of Vacancies and Interstitials in Helium-4}
\author{Ping Nang Ma}
\email{tamama@hku.hk , pingnang@phys.ethz.ch}
\affiliation{Center of Theoretical and Computational Physics, The University of Hong Kong, China}
\affiliation{Department of Physics, The University of Hong Kong, China}
\author{Lode Pollet}
\author{Matthias Troyer}
\affiliation{Theoretische Physik, Eidgen\"ossiche Technische Hochschule Z\"urich, 8093 Z\"urich, Switzerland}
\author{Fu Chun Zhang}
\affiliation{Center of Theoretical and Computational Physics, The University of Hong Kong, China}
\affiliation{Department of Physics, The University of Hong Kong, China}

\begin{abstract}
Motivated by experimental hints for supersolidity in Helium-4, we perform Monte Carlo simulations of vacancies and interstitials in a classical two-and three-dimensional Lennard-Jones solid. 
We confirm a strong binding energy of vacancies which is of the order of Lennard-Jones attraction.
This is reminiscent of what has been found for vacancies in Quantum Monte Carlo simulations. 
In addition, we find a strong attraction and large binding energy of interstitials in two-dimensional simulations.
This is mainly due to the formation of a pair of dislocations by clustering interstitials, in which minimizes the elastic deformation energy.
We interpret the results in light of the properties of Helium-4.
\end{abstract}

\maketitle

\section{Introduction}

A supersolid is a conjectured phase of matter which possesses seemingly contradicting properties of having spatial long range crystalline (positional and orientational) order as well as being superfluid at the same time. 
Potential evidence for such a supersolid phase in Helium-4 has been first detected as nonclassical rotational moment of rotational inertia (NCRI) in torsional oscillator experiments by Kim and Chan (KC) in 2004,\cite{KC1,KC2} more than thirty years after the first proposals by Andreev and Lifshitz \cite{AL} and Chester.\cite{Chester}
Based on their idea of a supersolid phase in a bosonic quantum crystal with vacancies (or interstitials), a number of  theoretical explanations have been proposed.\cite{Pederiva,Chaudhuri,Galli,Dai}

Although NCRI for Helium-4 has also been observed by other experimental groups,\cite{Rittner,Kondo,Penzev} it remains controversial whether if this is a supersolid phase or if the interpreted supersolid phase is simply a bulk equilibrium phenomenon.\cite{Rittner,Kondo,Penzev,Day,Sasaki}
Computer simulations have shown that the density matrix of Helium-4 crystal decays exponentially \cite{superglass,Clark} and vacancies in solid Helium-4 are gapped\cite{Lode}.
Non-equilibrium vacancies present in solid Helium-4 attract each other and phase separate, thus purging the Helium crystal of vacancies.\cite{Lode} 
This shows that the ground state of a single crystal Helium-4 is not a supersolid\cite{theorem}. It is remarkable that the solid with the strongest quantum properties known in nature behaves so classically.


Therefore, the understanding of the effective vacancy-vacancy interaction and interstitial-interstital interaction is crucial to our understanding of this phenomenon.
Here we set to carry out a classical Monte Carlo simulation in order to compare with the properties of Helium-4.
We find that a classical modeling of Helium yields vacancy and interstitial binding energy that are larger than that in quantum case.
Nonetheless, the phase remains essentially the same.

\section{Simulating a classical Lennard-Jones crystal}

For our classical Monte Carlo simulations we model the interaction between two atoms by a Lennard-Jones potential, where we fix units by choosing parameters suitable for Helium-4:\cite{Low Temperature Physics}
\begin{equation}
\phi(r) = 4\epsilon[({\frac{\sigma}{r}})^{12} - ({\frac{\sigma}{r}})^{6}]
\end{equation}
with $\sigma = 2.56\AA$ and $\epsilon/k = 10.2K$. 

We perform classical Monte Carlo simulation in the NPT ensemble of constant particle number, pressure, and temperature using two types of moves.
The first type is a local particle displacement: we propose to move a randomly chosen particle by a random distance in the interval $[0,\sigma)$ for the simulation of vacancies in a crystal, and by a random distance in the interval $[0,\sigma/4)$ for the simulation of a crystal with interstitials. The second type is a volume change, where we propose to vary the volume of the system by adjusting the length, width and height of the system individually within the range of $\pm 2.5\%$. 

Initially, we start from a perfect triangular lattice for our two-dimensional (2D) or from perfect hexagonally closed packed (hcp) lattice for our three-dimensional (3D) simulations. 
For a solid with vacancies, we randomly remove particles from the system, while
for a solid with interstitials, we randomly add particles to the system.
In our simulations, we define one sweep as $1000N^2$ trial particle moves (where N is the number of particles in the system), followed by one volume-changing move in each dimension. 
While quantum mechanically Helium-4 forms a crystal only at a pressure of above 25.46 bar, classically a crystal is already formed at zero pressure. 
Therefore, we use a pressure of only $P=1$ bar in our classical simulations. 
In fact, the exact value of pressure does not matter because we are only interested in qualitative features of a classical crystal rather than quantitative results for Helium-4.

We probe for attractive interactions and binding of vacancies and interstitials by calculating the binding energies and inverse compressibility.
The inverse compressibility can be measured as
\begin{equation}
\kappa ^{-1}= \frac{\partial ^2 E}{\partial N^2}
\end{equation}
where $N$ is the number of particles in the system. 
Thermodynamically, phase separation will occur when $\kappa^{-1} < 0$.
Numerically, the inverse compressibility is computed as
\begin{equation}
\kappa^{-1} = E(N+1) + E(N-1) - 2\,E(N).
\end{equation}

The binding energy of a system with $p$ vacancies can be expressed as
\begin{equation}
\label{BE_vac_eq}
E_B(N_0,p) = E(N_0-p) + (p-1)E(N_0) - p\,E(N_0-1)  \,\,\,\,\, ,
\end{equation}
and that of $p$ interstitials as
\begin{equation}
\label{BE_int_eq}
{\it E_B(N_0,p) = E(N_0+p) + (p-1)E(N_0) - p\,E(N_0+1)}  \,\,\,\,\, ,
\end{equation}
where $N_0$ is the number of particles of the commensurate crystal.

\section{Vacancies in a classical crystal}
For the simulation of vacancies we consider  systems with 1, 2 and 3 vacancies respectively. 
To find the ground state, we perform an annealing process. 
Due to the simple structure of energy space of vacancy systems, a rapid annealing schedule can be used. 
We start from a temperature of 1K and reduce it by a factor of 2 every 100 sweeps until we reach a final temperature of 0.01K. 

\begin{figure}
\includegraphics[width=8cm]{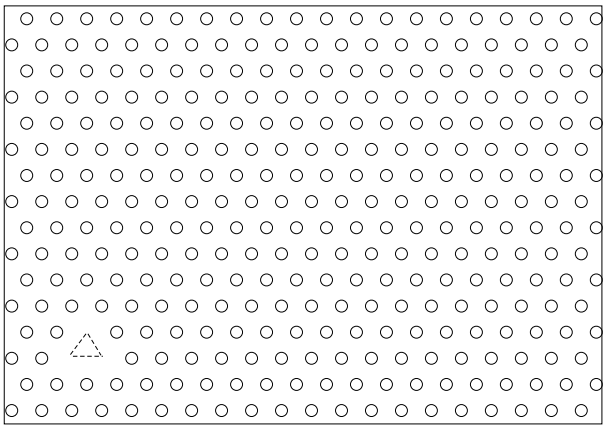}
\caption{Snapshot of a typical configuration, showing a stable vacancy cluster, $N_0=400$ with three vacancies, $T=0.01$K. Periodic boundary conditions are used.}
\label{vac}
\end{figure}

As shown in Fig. \ref{vac}, vacancies have a tendency to bind and form a vacancy-cluster. 
The three-vacancy cluster shown to be the ground state of the classical system is also the most likely configuration in Quantum Monte Carlo simulations by Ref. \onlinecite{Lode}.

\begin{figure}
\includegraphics[width=8cm]{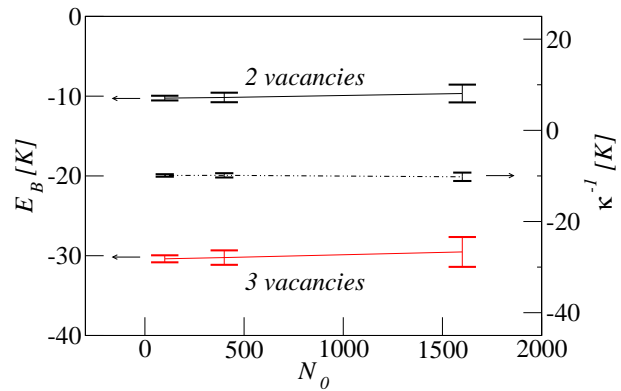}
\caption{Binding energy $E_B$ and inverse compressibility $\kappa ^{-1}$ of two and three vacancies in two dimensions at $T=0.01K$ as a function of the number  of particles in a commensurate solid $N_0$. The arrows indicate on which axis to read off the values of the data points.}
\label{vac_graph_1}
\end{figure}

\begin{figure}
\includegraphics[width=8cm]{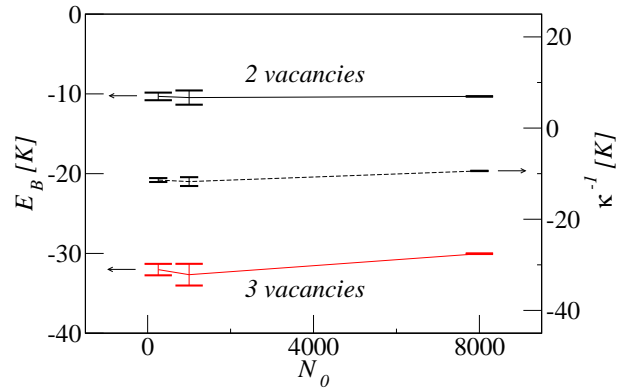}
\caption{
Binding energy $E_B$ and inverse compressibility $\kappa ^{-1}$ of two and three vacancies in three
 dimensions at $T=0.01K$ as a function of the number  of particles in a commensurate solid $N_0$.}
\label{vac_graph_2}
\end{figure}

In  Figs.  \ref{vac_graph_1} and  \ref{vac_graph_2}, we show two-vacancy binding energy, three-vacancy binding energy, as well as two-vacancy inverse compressibility at temperature $T=0.01$K and pressure $P=1$ bar for two and three dimensions.

\begin{figure}
\includegraphics[width=8cm]{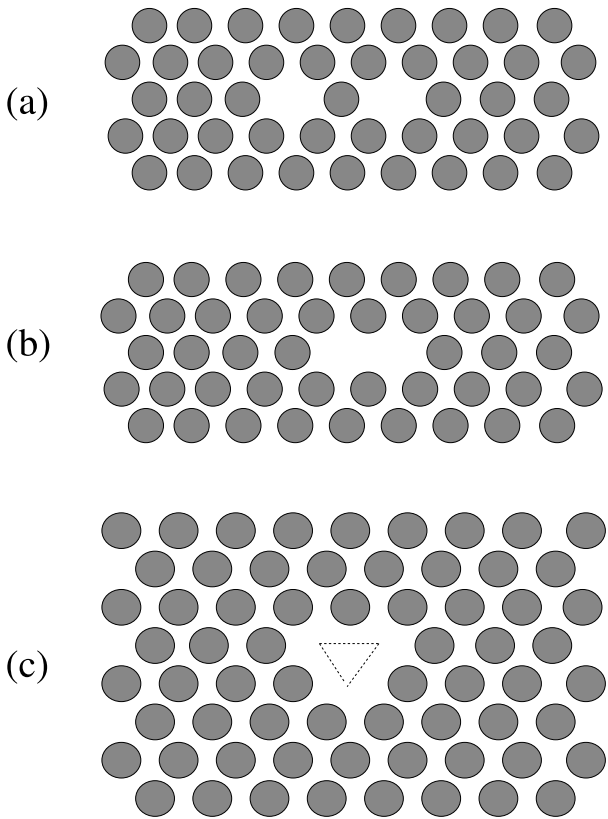}
\caption{The binding of vacancies: a) shows the configuration before and b) after the formation of a two-vacancy cluster. By binding vacancies the number of nearest neighbor atoms is increased by one and the bound state stablized.  c) The three vacancies form three pairs, indicated by the three dashed lines. As a result, a three-vacancy cluster is expected to have a binding energy three times than that of a two-vacancy cluster.}
\label{vac_explain}
\end{figure}

From the Lennard-Jones potential, the classical minimum interaction energy between two Helium atoms at zero temperature and zero pressure is exactly $\epsilon = -10.2$K. (This is slightly higher at finite pressure or temperature.)
Comparing Fig. \ref{vac_explain}a) and b), we realize that if two vacancies cluster, the system will be stabilized by the energy of one pair of atoms, which costs about -10K at finite temperature and pressure.
This argument shows that our simulation results for the binding energy of two vacancies in Fig. \ref{vac_graph_1} and \ref{vac_graph_2} are sensible.

For a three-vacancy cluster (Fig. \ref{vac_explain}c) there are three pairs of interaction and the binding energy is therefore approximately three times as much as that of a two-vacancy cluster.
Therefore, the value of -30K is once again expected. The clustering of vacancies can also be understood from minimizing the surface energy of a vacancy cluster. 
The surface in contact with the crystal is much lower for a vacancy cluster than for individual vacancies.

Similar effects have been seen in quantum Monte Carlo (QMC) simulations, where a binding energy of around one Kelvin for two vacancies and of several Kelvin for three vacancies has been observed \cite{Lode,LodePrivate}. 
The substantial reduction of the binding energy is mainly due to quantum zero point effect. Although these values are much reduced, they are still sufficient to cause phase separation of vacancies.\cite{Lode} 
Qualitatively, the classical picture applies also to the quantum Helium-4 system.

\section{Interstitials in a classical crystal}
Unlike the case of vacancies, where it is relatively easy to locate the global minimum, the interstitial system possesses many meta-stable local minima. 
To search for the ground state, we have to perform a stimulated tempering simulation.
Instead of using the much more involved conventional parallel tempering approach\cite{Helmut,Trebst,Trebst2,WL}, we use a simpler method that works well in our interstitial case.

\begin{figure}
\includegraphics[width=8cm]{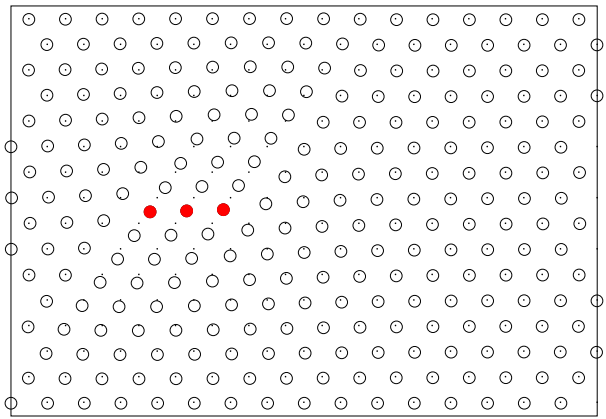}
\caption{A typical configuration of three interstitials, $N_0=256$ with three interstitials, $T=10^{-7}$K. Open circles represent the position of the He-4 particles; dots represent an imaginary perfect lattice; and solid circles represent the interstitials. To select the interstitials, we first superpose the actual configuration onto a virtual perfect consumerate lattice. After which, we pair up each lattice point with the closest actual particle in the configuration. Finally, we remove all these pairs and what remains will defined as the interstitials. The final aspect ratio of the configuration is 0.889 (c.f. 0.867 of a NVT configuration). Here, we want to emphasize the importance of not fixing the aspect ratio, because one will never obtain the lowest energy configuration once this aspect ratio is fixed.}
\label{int}
\end{figure}

The updates were identical to the case of vacancies, except that one row of atoms was fixed to prevent the crystal from rotating. 
Otherwise, the interstitials will be removed when the system chooses a different orientation by rotating.
From an initial temperature of 0.1K the temperature is decreased in an annealing procedure, taking care that the energy does not drop by more than 0.2K in 100 sweeps. 
Repeated simulations (starting from different initial conditions) have been carried out.
More than $80\%$ of these simulations yield the same lowest energy configuration, which we can safely call the global minimum. Our results again show binding of interstitials. A typical configuration is shown in Fig. \ref{int}.

We observe that at low temperatures the interstitials will be individually inserted into neighboring rows, creating a pair of dislocations, connected by a line of extra particles (see Fig. \ref{int}).  This is in contrast to vacancies which have a tendency to cluster into compact vacancy clusters.

\begin{figure}
\includegraphics[width=8cm]{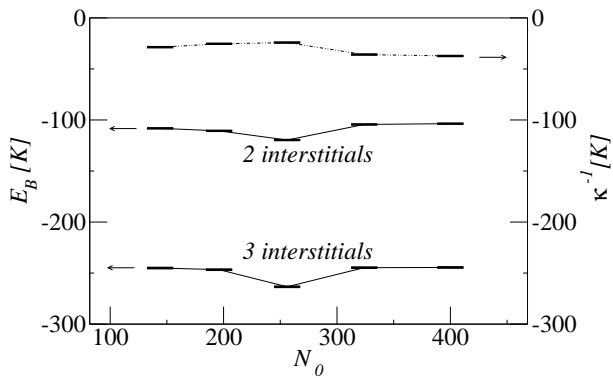}
\caption{Binding energy $E_B$ and inverse compressibility $\kappa ^{-1}$ of 
 of two and three interstitials in two dimensions s a function of the number  of particles in a commensurate solid $N_0$.}
\label{inter_graph}
\end{figure}

The binding energy and inverse compressibility are shown in Fig. \ref{inter_graph}.
Surprisingly we find even larger binding energies for interstitials than for vacancies. 
Interstitials cost more energy because they significantly distort the crystal, and much of this extra energy can be recovered by the binding of interstitials into a line connecting two dislocations as seen in Fig. \ref{int}.

\section{Conclusions}

Our classical Monte Carlo simulations of vacancies and interstitials in a classical solid show strong attraction and binding of vacancies and interstitials in a Lennard-Jones crystal. This is highly detrimental to the formation of a supersolid in the quantum mechanical case, where vacancies and interstitials are delocalized. Similar full quantum simulations of vacancies in Helium-4 show that while the binding is strongly reduced by quantum fluctuations, it is still large enough to cause vacancies to be expelled from the crystal instead of forming a supersolid.  Although the behavior of interstitials in solid Helium has not been studied in detail yet (apart from a lattice modeling~\cite{Manousakis}), it is established that they cost a finite energy gap $\Delta = (22.8 \pm 0.7) K$~\cite{Lode} to create. The strong binding seen in the classical case combined with the larger effective mass for interstitials indicates that the strong attraction will remain in the quantum case.

The ALPS libraries \cite{ALPS} were used for parallelization and error evaluation.  
The simulations were performed on the Hreidar cluster of ETH Z\"urich and the hpcpower cluster of HKU.
We treasure useful discussions with J. Wang, G. H. Chen, S. Q. Shen, and P. Corboz.
Financial support from RGC of HKSAR is also greatly acknowledged.

\end{document}